\documentclass[a4paper,twoside,reqno,10pt]{article}
\usepackage{amssymb,amsmath,amsthm}
\usepackage{graphicx}
\usepackage{times}\usepackage{color}
\usepackage{natbib}
\leftmargini=\baselineskip
\newtheoremstyle{localthm}
	{7pt} % space above
	{7pt} % space below
	{\sl} % Body font
	{} % Indent amount
	{\bf} % Theorem head font
	{{\rm.}} % Punctuation after theorem head
	{.7em} % Space after theorem head
	{} % Theorem head spec ?

\theoremstyle{localthm}
\newtheoremstyle{localrem}
	{5pt} % space above
	{5pt} % space below
	{\rm} % Body font
	{} % Indent amount
	{\bf} % Theorem head font
	{{\rm.}} % Punctuation after theorem head
	{.7em} % Space after theorem head
	{} % Theorem head spec ?

\theoremstyle{localrem}

\def\rss{\mathrm{rss}}
\def\RSS{\mathrm{RSS}}

\def\rss{\mathrm{rss}}
\def\RSS{\mathrm{RSS}}

\def\bs{\boldsymbol}

\def\veps{\bs{\varepsilon}}

\def\x{\bs{x}}

\def\y{\bs{y}}
\def\Y{\bs{Y}}

\def\Z{\bs{Z}}
\def\P{\bs{P}}

\begin{document}
%\addtolength{\baselineskip}{0.4\baselineskip}
% %==================================================================
\begin{center}
{\large An Approximation Based Theory to Linear Regression.}

\quad\\
Laurie Davies\footnote{ Faculty of Mathematics, University of
  Duisburg-Essen, 45117 Essen, Federal Republic of
  Germany. e-mail:laurie.davies@uni-due.de} 
\end{center}
\quad\\
\begin{abstract}
The goal of this paper is to provide a theory  linear regression based entirely on approximations. It will be argued that the standard linear regression model based theory whether frequentist or Bayesian has failed and that this failure is due to an 'assumed (revealed?) truth' (John Tukey) attitude to the models.  This is reflected in the language of statistical inference which involves a concept of truth, for example efficiency, consistency  and hypothesis testing. The motivation behind this paper was to remove the word `true' from the theory and practice of linear regression  and to replace it by approximation. The approximations considered are the least squares approximations. An approximation is called valid if it contains no irrelevant covariates. This is operationalized using the concept of a Gaussian P-value which is the probability that pure Gaussian noise is better in term of least squares than the covariate. The precise definition given in the paper is intuitive and requires only four simple equations. Given this a valid approximation is one where all the Gaussian P-values are less than a threshold $p0$ specified by the statistician, in this paper with the default value 0.01. This approximations approach is not only much simpler it is overwhelmingly better than the standard model based approach. This will be demonstrated using six real data sets, four from high dimensional regression and two from vector autoregression. Both the  simplicity and the superiority of Gaussian P-values derive from their universal exactness and validity. This is in complete contrast to standard F P-values which are valid only for carefully designed simulations.

 The paper contains excerpts from an unpublished paper by John Tukey entitled `Issues relevant to an honest account of data-based inference partially in the light of Laurie Davies's paper'.

\end{abstract}

%\tableofcontents

\section{Introduction}
John von Neumann (\cite{wrks47})
\begin{quotation}
I think that it [mathematics]  is a relatively good approximation to truth - which is much too complicated to allow anything but approximations 
\end{quotation}

The approach to linear regression developed in this paper is neither frequentist or Bayesian. There are no expected values, no prior distributions, no posterior distributions, there is no hypothesis testing,   no concept of efficiency, no concept of consistency,  no likelihood,  no need for post selection inference and so on. All of these involve a concept of `truth'. The most cited paper of  JRSS B, \cite{BENJHOCH95}, is based entirely on hypotheses being either `true' or  `not-true'. The author once asked a philosopher of statistics about this, the reply was `it is just a manner of speaking'.  It is an attitude unworthy of statistics. 

The use of truth derives from modelling together with what Tukey  (\cite{TUK93B}) calls `assumed (revealed?) truth' (note the word `honest' in the title). That is, the statistician behaves as if the model were true knowing perfectly well that `all model are false' (George Box). The result is the adaption of the language and concepts, including optimality considerations, which are only valid if the data were in fact generated as specified by the model which they were not. 

In this paper we adapt part of the von Neumann citation to linear regression `which is much too complicated to allow anything but approximations'. The approach to linear regression is based on two separate ideas, replacement of a model as  by approximations and the replacement of standard F P-values by Gaussian P-values. As stated in the Abstract a valid approximation is one all of whose Gaussian P-values do not exceed a threshold $p0$, here and and in the R package {\it gausscov} with default value $p0=0.01$. If the number $q$ of available covariates is large this translates into all standard F P-values being at most $0.01k/q$  where $k$ is the size of the approximation. The exact expression is derived below. 

We give two examples. The first is the riboflavin data with dimensions $(71,4088$  (\cite{DEBUEMEME15}). The two approximations below were obtained using the Gaussian covariate method 
\begin{eqnarray}\label{equ:ex1}
0.186\text{: } 71, 434, 1172, 1849, 1913, 2564, 2694, 3055,  3973,  4003.\nonumber\\
0.200\text{: } 314, 624, 1091, 1342, 1827, 1869, 1987, 2172,  3479,  3978,  4006\nonumber
\end{eqnarray}
where 0.186 and 0.200 are the standard deviation of the residuals. Both are valid as all the standard F P-values are all less than  $0.01k/q\sim 2.446$e-05 as the reader can check by running simple linear regressions.  The correlation between the two approximations is 0.962. Figure~\ref{fig:riboflavin} shows the dependent variable, the first approximation and the residuals. Section~\ref{R2} gives the R code for reproducing the approximations. The two approximations can be converted into two approximate models by adding Gaussian white noise with $\sigma$ equal to the respective standard deviations of the residuals. The two models are beyond the reach of model based procedures. The author know of no model based procedure which can get even close to these approximate models. Furthermore they cannot be evaluated using standard statistical inference and its ``true, not-true'' dichotomy, \cite{BENJHOCH95}.

The second example is vector autoregression using USA  economic data from
{\footnotesize
\begin{verbatim}
https://research.stlouisfed.org/econ/mccracken/fred-databases/
\end{verbatim}
}
\noindent
The original data set has dimensions $(256,233)$. All variables with an `NA'were removed giving  a reduced data set with dimensions $(256,182)$ . The dependent variable used here is the first which is the USA GDP . The covariates are the lags 1:10 of all the 182 variables giving 1821 covariates in all including the intercept. The approximation 
\begin{equation}\label{equ:ex2}
67.02\text{: } 11, 71 ,398,  431, 437, 445, 621, 701, 937, 1301, 1761.
\end{equation}
is valid as all F P-values are less than $6.044$e-05. Section~\ref{R1} gives the R code for reproducing the results. Figure~\ref{fig:vardata} shows the dependent variable, the approximation and the residuals. A model can be obtained by modelling the residuals. This is done in  Section~\ref{R1} and results in the model $veps(t)=N(0,67^2)$.

The importance of this example is that  the covariates are functions of the dependent variable, are non-stationary and yet they were analysed in exactly the same manner as the riboflavin data with fixed covariates. This is only possible because the Gaussian P-values are valid in both cases, for fixed covariates, for lagged covariates and indeed for all covariates in complete contrast to F P-values. It is this lack of validity for F P-values which has given rise to a large specialized literature for vector autoregression. Gaussian P-values render it irrelevant.

Section~\ref{sec:gau_P_val0} gives a short introduction to Gaussian P-values and approximations in linear regression. Section~\ref{sec:fncts} describes the two most important functions based on Gaussian P-values, {\it f1st} and {\it f3st}. Much of the paper is concerned with analysisng real data sets: the riboflavin, leukemia, the osteoarthritis, the Boston housing, the sunspot  and USA economic data sets are considered in Section~\ref{sec:fur_exam}. In Section~\ref{sec:graphs} the Gaussian P-value method is applied to the the construction of directed and undirected graphs. A first attempt at statistical inference based on up to hundreds of approximations is made in Section~\ref{sec:eval_approx}. Section~\ref{sec:mode_based_selecion} compares several model based procedures using the same data sets. The design of simulations  is considered in Section~\ref{sec:simulations}. It is argued that the simulations are confirmatory in that they are designed to give the desired sparse answer and bear little resemblance to real data.   % A different approach to approximation is considered in Section~\ref{sec:submodels} where the behaviour of submodels is considered without assuming a model for the data.  
Section~\ref{sec:trth_st_appr} is concerned with the role of truth in statistical analysis and includes along excerpt from \cite{TUK93B} which has never been published. 

\section{Gaussian P-values and approximations} \label{sec:gau_P_val0}

\subsection{Gaussian P-values}
The idea of Gaussian P-values is very simple. The simplicity  is shown by the fact that the relevant theory is covered by the next four equations. It could all have been done sixty or more years ago.  Given a dependent variable $\y$ of size $n$ and a subset ${\mathcal S}$ of covariates the Gaussian P-value $P_{G,\mathcal S}(\x_i)$  of $\x_i \in {\mathcal S}$ is obtained by replacing $\x_i$ by a Gaussian covariate $\Z$ consisting of i.i.d. $N(0,1)$ random variables to give a random sum of squared residuals $\RSS_{\Z\cup {\mathcal S}\setminus \x_i}$. 

The Gaussian P-value of $\x_i$ is defined as
\begin{equation} \label{equ:rss}
P_{G,{\mathcal S}}(\x_i)=\P( \RSS_{\Z\cup {\mathcal S}\setminus \x_i}<\rss_{\mathcal S}),
\end{equation}
that is, the probability that $\Z$ is better than $\x_i$. The distribution of $\RSS_{\Z\cup {\mathcal S}\setminus \x_i}/\rss_{{\mathcal S}\setminus \x_i}$ is given by 
\begin{equation} \label{equ:distRSS}
 \RSS_{\Z\cup {\mathcal S}\setminus \x_i}/\rss_{{\mathcal S}\setminus \x_i}\sim   \text{Beta}((n-k)/2,1/2)
\end{equation}
where $k$ is the size of ${\mathcal S}$ and $\rss_{{\mathcal S}\setminus \x_i}$ denotes the sum of squared residuals when $\x_i$ is excluded (see  \cite{DAVDUEM22} for a proof due to Lutz D\"umbgen).  This is the most important equation in the theory of Gaussian P-values, namely it gives an simple explicit expression for the distribution of $ \RSS_{\Z\cup {\mathcal S}\setminus \x_i}/\rss_{{\mathcal S}\setminus \x_i}$. Furthermore it is universally valid, that is,  it is valid for all data and every subset ${\mathcal S}$. It is valid even if the covariates are functions of the dependent variable $\y$ as in  vector autoregression. Without it there would be no point in Gaussian P-values.

From (\ref{equ:rss}) and (\ref{equ:distRSS}) it follows
\begin{equation} \label{equ:rss2}
P_{G,{\mathcal S}}(\x_i)=\text {Beta}_{(n-k)/2,1/2}(rss_{\mathcal S}/\rss_{{\mathcal S}\setminus \x_i}).
\end{equation}
which inherits the universally validity of (\ref{equ:distRSS}). 

The number $q$ of covariates can be taken into account by defining 
\begin{equation} \label{equ:rssq2}
P_{G,{\mathcal S}}(\x_i)=1-(1-P_{G,{\mathcal S}}(\x_i))^{q-k+1} 
 \end{equation}
where the right hand $P_{G,{\mathcal S}}(\x_i)$ is given by (\ref{equ:rss2}). In  (\ref{equ:rss}) the covariate $\x_i$ is only compared with a single Gaussian covariate. If given ${\mathcal S}$ all the remaining $q-k$ covariates are replaced by i.i.d. Gaussian covariates then the statistician will not want to include any of these in a selected subset. To avoid this  $\x_i$ is compared to the best of the now $q-k+1$ Gaussian covariates which leads to (\ref{equ:rssq2}). 

The Gaussian P-value (\ref{equ:rss}) and the standard F P-value are equal,
\begin{equation} \label{equ:GP}
P_{G,{\mathcal S}}(\x_i)=P_{F,{\mathcal S}}(\x_i).
\end{equation}
This follows from (\ref{equ:rss2}). A proof based on Cochran's theorem has been given by Joe Whittaker. Both the Gaussian P-value and the F P-value are functions of the statistic $rss_{\mathcal S}/\rss_{{\mathcal S}\setminus \x_i}$. In the case of Gaussian P-values the Beta distribution in (\ref{equ:distRSS}) derives from the Gaussian covariate $\Z$ whereas the corresponding distribution of the standard F statistic derives from the Gaussian error term $\veps$ in the model. The consequence is that the former is universally valid, the latter only for specially designed simulations.

 As F P-values are all derive from the same error term $\veps$ there is no equation corresponding to (\ref{equ:rssq2}) for F P-values but from (\ref{equ:GP})  we have
\begin{equation} \label{equ:rssq3}
P_{G,{\mathcal S}}(\x_i)=1-(1-P_{F,{\mathcal S}}(\x_i))^{q-k+1}.
 \end{equation}
which allows Gaussian P-values to be directly calculated from standard F P-values. 

A full account of  Gaussian P-values is given in \cite{DAVDUEM22}, see also \cite{DUMDAV23}.

\subsection{Approximations}

Given data $(\y,\x)$ and a subset ${\mathcal S}$  of covariates the approximation based on ${\mathcal S}$ is is the linear combinations of the covariates in ${\mathcal S}$ using the least squares values of the coefficients when $\y$ is regressed on the subset ${\mathcal S}$ of covariates. No assumptions are made about the data, the approximation is to the data as given, no matter what this is.

To base statistical analysis on approximations runs counter to the whole tradition of statistics and little interest in it has been shown. One exception was John Tukey, see \cite{TUK93B}, \cite{TUK93C} and \cite{TUK93D}. These papers have never been published but Brillinger briefly mentions \cite{TUK93B} in \cite{BRIL02A}. Tukey's papers were a reply to a first version of \cite{DAV95} which was a first attempt to think in terms of approximation. Another exception is Peter Huber, see Chapter 5 of \cite{HUB11}. The following papers were written in this sense \cite{DAVKOV01}, \cite{DAV08} (written with the encouragement of Peter Hall), \cite{DAV14} and \cite{DAV18a}. 

\section{Functions based on Gaussian P-values} \label{sec:fncts}

\subsection{The function {\it f1st}} 

The basic selection function of {\it gausscov} is {\it f1st}. It is a greedy fast forward selection procedure and returns either no subset or just one. Given a subset ${\mathcal S}$ of size $k$ already selected the next candidate for selection is the best of the remaining $q-k$ covariates $\x_b$. Its Gaussian P-value is given by
\begin{equation} \label{equ:eq3}
\P_{G,{\mathcal S}}(\x_b)=1-(1-P_{F,{\mathcal S}\cup \x_b}(\x_b))^{q-k}
\end{equation}
which differs from (\ref{equ:rssq2}) by 1 as $\x_b\notin {\mathcal S}$. If this is less than $p0$ $\x_b$ is selected and the procedure continues. If not the procedure terminates with ${\mathcal S}$. There is no guarantee that ${\mathcal S}$ is valid. If the size $k$  of ${\mathcal S}$ is less than parameter {\it mx} with default value $mx=21$ an all subset search is conducted {\it sub = TRUE} which involves all $2^k$ subsets. The best valid subset is returned. If $k$ exceeds {\it mx}  ${\mathcal S}$ is returned and may not be valid. 

It can happen that the very first P-value exceeds $p0$ in which case no subset is returned. This is the case for the dental data of \cite{SEHTUK01} although there are valid subsets. This can be avoided by choosing a value for {\it kmn}. If this is done the first {\it kmn} covariates are selected irrespective of their P-values. After this the  greedy fast forward selection procedure continues as described. Even if {\it f1st} returns a subset it is often of advantage to set for example $kmn=10$ or $kmn=15$. This is the case for the leukemia data. With {\it kmn=0} just three covariates are selected, with {\it kmn=10} an additional three are selected, see Section~\ref{sec:leukemia1}.

The inclusion of an intercept is accomplished by the default setting of {\it inr=TRUE} and {\it xinr=FALSE}. If an intercept is already included set  {\it inr=FALSE} and {\it xinr=TRUE} but in this case it must be the last covariate otherwise it will be treated as any other covariate. The last parameter is {\it kex} which excludes covariates. It is an integral part of {\it f3st}.

\subsection{The function {\it f2st}} 
The function {\it f2st} was the first attempt at selecting several approximations. Given {\it kmn} {\it f1st(y,x,kmn)} is run and results in the covariates  $(\x_{i1},\ldots \x_{ik_1})$. {\it f1st(y,x,kmn)} is now run again but excluding the covariates  $(\x_{i1},\ldots \x_{ik_1})$, 
\begin{verbatim}
f1st(y,x,kmn,kex=(x_{i1},...,x_{ik_1}))
\end{verbatim}
This results in the covariates $(\x_{ik_1+1},\ldots,\x_{ik_2})$ which are also eliminated in the second step
\begin{verbatim}
f1st(y,x,kmn,kex=(x_{i1},...,x_{ik_2}))
\end{verbatim}
and the process continues until no further covariates are chosen. The approximation are based on disjoint subsets of covariates and are not very good as many relevant covariates are eliminated at an early stage.

\subsection{The function {\it f3st}} 
We demonstrate {\it f3st} using the riboflavin data. We start with {\it f1st(y,x)} which results in the covariates  (73, 2034, 2564, 4003). We eliminate each one in turn starting with 73. That is we run {\it f1st(y,x,kex=73} which results in the covariates (143,1278,2564,4003). We do this for 2034, 2564 and 4003 giving now in all five approximations. This is the result of {\it f3st(y,x,m=1)}. Now take the approximation (143,1278,2564,4003) obtained by excluding 73. Again we exclude each covariate in turn but also excludes 73 as well. That is we run   {\it f1st(y,x,kex=(73,143))}, {\it f1st(y,x,kex=(73,1278))}, .... ,  {\it f1st(y,x,kex=(4003,4006))}. This is the results of  {\it f3st(y,x,m=2)}. This continues in the same manner for larger values of {\it m}.

\subsubsection{Two examples using {\it f3st}} 
{\bf riboflavin data}\\
We apply {\it f3st} to the riboflavin data. With the parameter values {\it f3st(y,x,m=1,kmn =15)} five approximations are returned the best two of which are the first two lines of Table~\ref{tab:ribo1}. The second line is the result of {\it f1st(y,x,kmn =15)}. A more intensive search with {\it f3st(y,x,m=6,kmn =15)} returns 243 valid approximations in about 30 minutes the best two of which are given in the last two lines; see Section~\ref{R2} for the R code.

{\footnotesize
\begin{table}[ht]
\begin{center} 
\begin{tabular}{ll}
$sd$& Selected covariates\\
\hline
0.296&   73 1131 1278 2140 2564 4006\\
0.345&  73 2034 2564 4003 \\
0.185& 71 434 1172 1849 1913 2564 2694 3055  3973  4003 \\
0.200& 314  624 1091 1342 1827 1869 1987 2172  3479  3978  4006\\
\end{tabular}
\caption{The riboflavin data: the first two approximations are the best out of four running {\it f3st(y,x,m=1,kmn=15)} , time about 1.5 seconds. The the last two are the best out of 243 running {\it f3st(y,x,m=6,kmn=15)} and are based on 243 covariates , time about 30 minutes.\label{tab:ribo1}}
\end{center}
\end{table}
}

The reader can compare this with the analyses of the riboflavin data given in \cite{BKL14}, \cite{DEBUEMEME15} and {\cite{LOCK17} which illustrate the difficulties with the modelling approach.\\

\noindent
{\bf  USA  economic data}\\
These are ordered as follows. The first 10 are the lags 1:10 of the first variable, the second 10 the lags 1:10 of the second variable and so on. 

Table~\ref{tab:vardat1} gives the best two approximations out of 11 using {\it f3st(y,x,m=1,kmn=16)}. The R procedure is given in Section~\ref{R1} and takes about 30 seconds.  Figure~\ref{fig:vardata} shows the results for the first approximation. It is clear that the data are highly non-stationary. This time the residuals $\veps$ are correlated 
\[\veps(t)=1.0578+0.223\veps(t-1)+\delta(t)\]
with $\delta(t)$ i.i.d. $N(0,65.3^2)$. See Section~\ref{R1} for the R code.
{\footnotesize
\begin{table}[bt] 
\begin{center}
\begin{tabular}{ll}
$sd$& Selected covariates\\
\hline
 66.9&  11   71  398  431  437  445  621  701   937  1301  1761  \\
67.9&     18   71   73  142  437  445  621   701   931  1301  1361 1761
\\
\end{tabular}
\caption{The USA economics data: the two best approximations out of 12 for the USA GDP economic data, time about 70 seconds. The standard deviation of the residuals $sd$ followed by the chosen covariates. \label{tab:vardat1}}
\end{center}
\end{table}
}
We note that {\it f3st} again returns multiple approximations.

\section{Further examples: the leukemia, the osteoarthritis, the Boston housing and the sunspot data sets}\label{sec:fur_exam}

\subsection{The leukemia data}\label{sec:leukemia1}
For the data see \cite{GOLETAL99}. The dimensions are $(72,3572)$. Running {\it f1st(y,x)} returns the three covariates 1182, 1219 and 288. Running {\it f1st(y,x,kmn=10)} returns the three additional covariates 183, 3038 and 2558, see Section~\ref{R1.5}. Running {\it f3st(y,x,m=4,kmn=15)} yields 58 subsets of which the best two are
{\footnotesize
\begin{table}[ht]
\begin{center} 
\begin{tabular}{ll}
$sd$& Selected covariates\\
\hline
 0.036&  29  157  346 1157 1652 1674 2399 2588  2888  3476  3539\\
 0.042& 612  926  979 1174 1536 1652 1865 1946  1949  2208  2389 2664  2720  3098  3308\\
\end{tabular}
\caption{The leukemia data: the two best approximations out of 58, time about three minutes \label{tab:leukemia}}
\end{center}
\end{table}
}

\subsection{The osteoarthritis data}
This data was analysed in \cite{COXBATT17}. The dimensions are $(129,48802)$. Running {\it f3st(y,x,m=4,kmn=15} yields 34 subsets of which the best two are

{\footnotesize
\begin{table}[ht]
\begin{center} 
\begin{tabular}{ll}
$sd$& Selected covariates\\
\hline
0.129 & 3630 13329 14724 17062 18758 29522 43770 44902 48025\\
 0.141&  3630  8034 11536 14724 17062 17125 29522 44902 \\
\end{tabular}
\caption{The osteoarthritis data: the two best approximations out of 34, time about two minutes \label{tab:osteo}}
\end{center}
\end{table}
}

\subsection{The Boston housing data 1}
We run {\it f1st(y,x,kmn=13)}. 11 of the 13 covariates are selected, the standard deviation of the residuals is 4.68. See Section~\ref{R31} for the R code.
{\footnotesize
\begin{table}[ht]
\begin{center} 
\begin{tabular}{cccc}
Selected covariates&$\beta$ values&Gaussian P-values&F P-values\\
\hline
  1 & -0.108413345 &3.028253e-03 &1.010438e-03\\
  2 &  0.045844929 &2.261121e-03 &7.542759e-04\\
 4&   2.718716303& 4.647190e-03& 1.551469e-03\\
   5&  -17.376023429& 3.628235e-06& 1.209413e-06\\
  6&   3.801578840& 8.669336e-19& 2.889779e-19\\
   8 & -1.492711460& 2.051113e-14& 6.837043e-15\\
    9 &  0.299608454& 8.990371e-06& 2.996799e-06\\
   10&  -0.011777973& 1.563456e-03& 5.214237e-04\\
   11 & -0.946524570& 2.770519e-12& 9.235063e-13\\
 12&   0.009290845& 1.668794e-03& 5.565743e-04\\
13&  -0.522553457& 6.421758e-25& 2.140586e-25\\
    0 & 36.341145004 &2.727265e-12& 2.727265e-12\\
\end{tabular}
\caption{The Boston housing data run with {\it f1st(y,x,kmn=13)}:  \label{tab:bost0}}
\end{center}
\end{table}
}

\subsection{The Boston housing data 2}
The third  example is high dimensional regression using the Boston housing data. The covariates  are all interactions of order eight and less of the original 13 covariates. The dimensions are $(506,203490)$. A full Bayesian approach to the data would requires  the covariance matrix of the covariates of size $10^{10}$. Not surprisingly it fails. The R procedure is given in Section~\ref{R3}. It  uses {\it f3st(y,x,m=1,kmn=15} and takes about 90 seconds. The results are given in Table~\ref{tab:bost1}. By far the he most relevant interaction is  191110,  $bostint[[2]][191110,]=(6, 6, 6, 6, 12,  0,  0, 0) =boston[,6]^4boston[,12]$ with a Gaussian P-value of $5.074$e-160.
{\footnotesize
\begin{table}[ht]
\begin{center} 
\begin{tabular}{ll}
$sd$& Selected covariates\\
\hline
2.978&20184  21876  66089 128946 137414 153647 187466 190771 190865 191110\\
& 202995 203120 203287\\
2.981&20184  21878  66466 150307 154607 181203 182144 185492 190689 190771\\
& 191110 197058\\ 
 \end{tabular}
\caption{The Boston data: the two best approximations out of 69, time about 18 minutes.  \label{tab:bost1}}
\end{center}
\end{table}
}

\subsection{The Boston housing data 3}
In the third  example for the Boston housing data are all interactions of order eight of the transformation $\x_i \rightarrow \lg(1+\x_i)$ of the original 13 covariates.  The results are given in Table~\ref{tab:bost2}. 
{\footnotesize
\begin{table}[ht]
\begin{center} 
\begin{tabular}{ll}
$sd$& Selected covariates\\
\hline
2.95& 320  6074  6619  21390  61255 156787 157766 183308 185175 190627 193298 202668\\
2.98&1403  1503  2251   6619 129151 153683 183308 185175 190627 190881 196725 201607\\

\end{tabular}
\caption{The Boston data: the two best approximations out of 13, time about 2 minutes.  \label{tab:bost2}}
\end{center}
\end{table}
}

\subsection{The Boston housing data 4}
This time the interactions of the last two examples are combined to give dimensions $(506,406980)$. This was just to see how well {\it f3st} could cope with this order of covariates. 11 approximations were returned in about 3 minutes.  The results are given in Table~\ref{tab:bost3}. 
{\footnotesize
\begin{table}[ht]
\begin{center} 
\begin{tabular}{ll}
$sd$& Selected covariates\\
\hline
3.12& 20184 148950 179484 184537 190660 190697 190781 203810 388665 392703 406934\\
3.14&20184  71364 182854 190696 190781 191074 191211 193720 203810 390145 406934\\

\end{tabular}
\caption{The Boston data: the two best approximations out of 11, time about 2 minutes.  \label{tab:bost3}}
\end{center}
\end{table}
}

\subsection{The sunspot data}\label{sec:sunspot}
The data are of size 3253 and give the average number of sunspots each month from January 1749 to January 2020. They are available from the Royal Observatory of Belgium

{\footnotesize
\begin{verbatim}
 http://www.sidc.be/silso/
\end{verbatim}
}
\noindent
\subsubsection{Autoregression} 
See Section~\ref{R41}

\noindent
We allow for a maximum lag of 150 and use {\it fist} which is sufficient for our purposes. It yields the seven lags (1, 2, 4, 6, 9, 27, 111) with a constant 3.67 and coefficients 
\[0.540, 0.124, 0.130, 0.707, 0.113, -0.080, 0.05\]
The standard deviation of the residuals is 24.43.\\

The data shows that large $snspt(i)$ is associated with large$\vert res(i)\vert$  so we standardize the residuals by dividing by $snspt(i)+10$, $tres=res/(snspt+10)$. The factor 10 is somewhat arbitrary but the number of sunspots is sometimes zero. The autocovariance function {\it acf} indicates that the {\it tres} are correlated so we perform an autoregression with maximum lag 10. This results in the six lags $(2,3,4,5,7,8)$  with a constant -0.05 and the coefficients
\[0.06, 0.13, 0.07, 0.14, 0.07, 0.10\]
The residuals $delta$ are uncorrelated and can be reasonably approximates as a $2.8Beta(12,6.7)-1.75$ random variable.

\subsubsection{Non-parametric regression}
We now take a non-parametric approach. The function {\it  fgentrig(n,m)}  returns the values 
\[\sin(\pi ij/n),\cos(\pi ij/n), i=1,\ldots,n, j=1,\ldots,m\]
which are the covariates. For the sunspot data $n=3253$ and we set $m=1626$. {\it f1st} with $p0=0.001$ returns a valid approximation based 50 covariates. The choice $p0=0.001$ gives a smoother function by removing a high frequency covariate. The resulting function $snf$ has negative values: to obtain a function with only positive values we put $snf=20\exp(\log(1+snf/20))$, shown in  the upper panel of  Figure~\ref{fig:snspt_mod}. The standard deviation ion of the residuals is 24.97 with the outliers removed and  25.3 including the outliers as compared with 23.82 for the autoregressive approach. The residuals are standardized by dividing by {\it snf}, $rest(i)=res(i)/snf(i)$. The autocorrelation functions shows that the {\it rest} are correlated. To remove this we perform an autoregression with maximum lag 10: the relevant lags are (1,2,3,5). The residuals are uncorrelated but exhibit about 36 outliers which we remove using Hampel's 5.2 rule, \cite{HAM85}. The resulting residuals are approximately distributed as $0.5t_{10}^{+}+0.6t_2^{-}$ where $t_{10}^{+}$ is the postive part of a $t_{10}$  random variable and $t_2^{-}$ is the negative part of a $t_2$ random variable, see the lower panel of Figure~\ref{fig:snspt_mod}.

\section{Graphs} \label{sec:graphs}
There are several methods for calculating dependency graphs. The ones we compare here are the Gaussian covariate graph function {\it fgr1st}, {\it thav.glasso} (\cite{LASZ21}), {\it huge}  (\cite{JFLRLWLZ21}) and {\it glasso} (\cite{FRHATI19}).
 
We consider firstly a random graph generated as in \cite{MEIBUE06} but with the correction given in \cite{DAVDUEM22}. The undirected graph procedure {\it fgr1st}  of {\it gausscov} was used. The procedures {\it thav.glasso} and {\it glasso} have regularization parameters {\it C} and {\it rho} respectively with no indication as to how to choose them. They were chosen to equalize approximately the numbers of false positives and false negatives. The values used were {\it C=0.4} and {\it rho=0.236}. Table~\ref{tab:rnd_graph} gives the results. The second column gives the number of edges, the third the number of false positives $fp$, the fourth the number of false negatives $fn$ and the fifth  the time in seconds. 

The results for the covariates of the riboflavin data were as follows. The procedures {\it thav.glasso} and {\it glasso} were killed after one hour with no results. There may be values of the regularization parameters which work but the user is given no hint. The {\it huge} procedure took 35 seconds but returned zero edges. The Gaussian covariate procedure with the default values took 16 seconds and yielded a directed graph with 4491 edges and an undirected graph with 3882 edges.

\begin{table}[tbp]
\begin{center}
\begin{tabular}{rcccc}
\multicolumn{5}{c}{Random graph (1000,1000)}\\
\hline\\
method&no. edges&$fp$&$fn$&time\\
\hline\\
{\it fgr1st} &1831&8&0&44\\
{\it thav.glasso} &1776&218&265&90\\
{\it huge}&1830&26&19&30\\
{\it glasso}&1840&293&276&16\\
\end{tabular}
\caption{The results for one simulation of the random graph. \label{tab:rnd_graph}}
\end{center}
\end{table}  

\section{Evaluating approximations}\label{sec:eval_approx}
There has been no work on how to evaluate multiple approximations. Here a first attempt for the riboflavin data using the 237  approximations from f3st.  Figure~\ref{fig:log_pval_ribo} plots the all the $-log(gpvalue)$ values for each of the 190 covariates. Care must be taken as the plot does not separate the covariates 4002:4008. Table~\ref{tab:frequencies} gives the frequencies of the 32 covariates with a frequency exceeding five.

 \begin{table}[ht]
 \begin{center} 
{\footnotesize
 \begin{tabular}{cccccccccccccc}
\hline
 69&   73 &143 &144 & 624 & 792& 1069 &1131& 1278 & 1279 & 1303 & 1425&  1516 & 1603\\
        15 &  65 &  32  & 33  & 22&    7 &   6  & 18&   36  &  35 &    9  &   6  &  12  &   7\\
\hline
1640 & 1762 & 1849&  2034&  2116 & 2117&  2140 & 2271  &2564 &3054 & 3055 & 3514&3525&4002\\
6 &    8 &    9  &  19  &  11  &  39&  14  &  26&   125&     9   &  8 &    8&9&25\\
\hline
  4003 & 4004 & 4005 & 4006\\
48 &   98  &  11 &   37\\
\end{tabular}
}
\caption{The frequencies of all covariates with a frequency of at least six.  \label{tab:frequencies}}
\end{center}
\end{table}

Another possibility is to calculate an undirected  dependency graph. The one derived from the 190 covariates has 1152 edges and the author is not sufficiently converse with graph theory to be able to derive any useful information from it. However Figure~\ref{fig:dep_graph} gives one based on the 32 covariates of Table~\ref{tab:frequencies} and  has only 34 edges.

\section{Model based procedures}\label{sec:mode_based_selecion}
The five data sets riboflavin, leukemia, osteoarthritis, Boston housing, sunspot and the USA economics data are evaluated using the following model based procedures: {\it scalreg} for scaled sparse linear regression (\cite{SUN19}) , {\it SIS} for Sure Independence Screening (\cite{FFSS20}), {\it stability} for stability selection, {\it multi-split} for multiple splitting , {\it ridge-proj} for ridge regression, (the last three available from \cite{MDMMB21}) and {\it lm.spike} for BoomSpikeSlab (\cite{SCO22}). The results are compared with those from {\it f1st} with $kmn=15$ and {\it f3st} with $m,kmn=15$. 

\subsection{The riboflavin data}  \label{sec:ribofl}
{\bf A simulation}\\
In contrast to the simulations of \cite{HASTIBTIB20} this simulation uses real data, namely the riboflavin data. The  covariates are standardized, six are chosen at random $I_1,\ldots,I_5$  and the dependent variable given by
\[\Y=10\sum_{j=1}^6\x_{I_j}+\veps\]
where $\veps$ is standard white Gaussian noise. The goal is to identify the $I_j$. 

 The results of 50 simulations are given in Table~\ref{tab:1st_simulation}.
\begin{table}[h]
\begin{center}
\begin{tabular}{rcccccccccc}
Function&f1st&f3st&ridge&split&SIS&stability&scalreg&spike\\
\hline
No. correct&23&36&18&1&0&0&0&13\\
time&1&7&800&100&10&12&10&100\\
\end{tabular}
\caption{The number of correct identifications in 50 simulations. The time is the factor slower than {\it f1st} which took 0.35 seconds on average. \label{tab:1st_simulation}}
\end{center}
\end{table}

Table~\ref{tab:ribo2} gives the results of analysing  the riboflavin data itself using the functions specified above. 
{\footnotesize
\begin{table}[bt]
\begin{center}
\begin{tabular}{lrrll}
function&time&$sd$&covariates\\
\hline
f1st&0.33&0.345&4003,2564,73,2034\\
f3st&1.65&0.296&73,1131,1278,2140,2564,4006\\
ridge&300.00&\#&none\\
split&36.00&0.720&4003\\
SIS&2.30&0.466&1516,1303,4004\\
stability&2.05&\#&none\\
scalreg&0.93&0.557&1303,4003&\\
spike&41.60&0.364&73,2034,2564,4006\\
lasso.proj&6.46&\#&none\\
\end{tabular}
\caption{The results for the riboflavin data. \label{tab:ribo2}}
\end{center}
\end{table}
}

\subsection{The leukemia data}\label{sec:leukemia2}

\begin{table}[ht]
\begin{center}
\begin{tabular}{lrrll}
function&time&$sd$&no. covariates\\
\hline
f1st&0.30&0.113&6\\
f3st&2.28&0.079&7\\
ridge&\#&\#&0\\
split&42.10&0.248&1\\
SIS&7.99&0.162&3\\
stability&1.73&0.209&2\\
scalreg&0.32&0.201&3\\
spike&26.70&0.178&3\\
lasso.proj&2129.00&0.177&4\\
\end{tabular}
 \caption{The results for the leukemia data\label{tab:leukem3}}
\end{center}
\end{table}

\subsection{The osteoarthritis data}\label{sec:osteo2}

\begin{table}[bt]
\begin{center}
\begin{tabular}{lrrll}
function&time&$sd$&no. covariates\\
\hline
f1st&0.86&0.229&3\\
f3st&3.45&0.220&3\\
ridge&\#&\#&Error: cannot allocate vector of size 17.7 Gb\\
split&480.00&0.273&2\\
SIS&25&\#&0\\
stability&\#&\#&Error in x[, ii] : subscript out of bounds\\
scalreg&13.00&0.331&2\\
spike&\#&\#&Error: cannot allocate vector of size 17.7 Gb\\
lasso.proj&\#&\#&Killed after two hours with no result\\
\end{tabular}
 \caption{The results for the osteoarthritis data\label{tab:osteo3}}
\end{center}
\end{table}
\newpage

\subsection{The Boston housing data}
The covariates  are all interactions of order eight and less of the original 13 covariates. The dimensions are $(506,203490)$.   As can be seen from Table~\ref{tab:bost3} only  {\it stability} did not end with an error.
 
\begin{table}[ht]
\begin{center}
\begin{tabular}{lrrll}
function&time&$sd$&no. covariates\\
\hline
f1st&9.69&3.15&11\\
f3st&134.00&2.93&13\\
ridge&\#&\#&Error in svd(x) infinite or missing values in x : \\
split&\#&\#&The classical.fit function didn't return the correct \\
&&&number of p-values for the provided submodel\\
SIS&\#&\#&NA/NaN/Inf in 'x\\
stability&919.00&6.15&2\\
scalreg&\#&\#&Error: cannot allocate vector of size 6.1 Gb\\
spike&\#&\#&Error: cannot allocate vector of size 308.5 Gb\\
lasso.proj&\#&\#&NA/NaN/Inf in foreign function call (arg 5)\\
\end{tabular}
 \caption{The results for the Boston housing data, see Table~\ref{tab:bost1} \label{tab:bost3}}
\end{center}
\end{table}
%\newpage

\subsection{The sunspot data}
{\footnotesize
\begin{table}[h]
\begin{center}
\begin{tabular}{lrrll}
function&time&$sd$&covariates\\
\hline
f1st&0.02&25.1&1 2 4 6 9 27 111\\
ridge&1.18&24.8&1  2  4  6  9 25\\
split&39& 26.5&1  2  3  4  9 43\\
SIS&\#&1\#&no. selected $>120$, error vector size too large\\
stability&0.90&25.1&1,2,3,4,6\\
scalreg&\#&\#&selected  149 lags \\
spike&1.30&24.4&1   2   4   6   9  18 108\\
lasso.proj&100.00&24.7&1  2  3  4  6  9 18 24 25\\
\end{tabular}
\caption{The results for the sunspot data. \label{tab:sunspot2}}
\end{center}
\end{table}
}

\subsection{The USA economics data}
\newpage
\begin{table}[ht]
\begin{center}
\begin{tabular}{lrrll}
function&time&$sd$&no. covariates\\
\hline
f1st&1.34&73.9&10\\
f3st&13.10&68.9&13\\
ridge&36.10&426.0&1\\
split&*&*&error, no inside point\\
SIS&0.62&156.0&1\\
stability&2.05&\#&none\\
scalreg&3.00&0.0&246 +error, vector specified set too large\\
spike&18.70&74.8&10\\
lasso.proj&1797.00&529.0&1\\
\end{tabular}
\caption{The results for the USA economic data. See Table~\ref{tab:vardat1} for the results of a more comprehensive search.\label{tab:vardat2}}
\end{center}
\end{table}

\section{Model based high dimensional regression and simulations}\label{sec:mod_hd} \label{sec:simulations}

 In \cite{HASTIBTIB20} the authors write
\begin{quotation}
 Rather  this paper is about:The relative merits of the three (arguably) most canonical forms for sparse estimation in a linear model: $\ell_0,\ell_1$  and forward stepwise selection.
\end{quotation}
As argued in the Introduction model based procedures are truth orientated. This is fully reflected in simulation study of \cite{HASTIBTIB20}. The number of non-zero parameter values is small reflecting the imposed sparsity condition,  5 for the dimensions 100 and 1000 and additionally 10 for 1000. Furthermore the parameter values are chosen so that the truth stands out and can be detected. The different procedures are compared using discrepancy measures to compare their distances from the truth the best one being that closest to the truth. There is only one truth so this prioritizes thinking in terms of optimization, getting as close to the truth as possible. This explains the treatment of the best subset procedure in the paper. This explains the language for example  ' a bias-variance tradeoff' , a truth based concept and so on. 

This can be compared with the Gaussian covariate procedure {\it f3st}. This is specially constructed to produce many valid approximations and does so for all the data sets considered. One could introduce the idea of a best approximations, that valid approximation with the smallest sum of squared residuals but this would not invalidate other valid approximations. Take the riboflavin data and apply  {\it f3st(y,x,m=6,kmn =15}. It returns 232 valid approximations. The best two are given in Table~\ref{tab:ribo1}. What is the truth and how are these two approximations to be evaluated by measuring their distance from the truth? 

The simulations bear no resemblance to real data: all the high dimensional data sets in this paper return multiple approximations but would not do so in the simulations of  of  \cite{HASTIBTIB20}. How can simulations be designed so that they do bear some similarity to real data? Here is a suggestion. Take the mean of the two best approximations of  Table~\ref{tab:ribo1}. This is a linear function of the 21 covariates land and add pure Gaussian noise with a standard deviation of say 0.185. This is the dependent variable. The first 21 of the 1000 covariates are those given in Table~\ref{tab:ribo1}, the remaining 979 are standard Gaussian white noise. One simulation using  {\it f3st(y,x,m=6,kmn =15)} resulted in 38 valid approximations, the residuals of the best two having standard deviations of 0.316 and 0.323. No Gaussian covariates were selected. It would be interesting to know how the procedures considered in \cite{HASTIBTIB20} would perform and how their performance would be evaluated.

\section{Truth, statistics and approximation}\label{sec:trth_st_appr}

\subsection{Tukey on Approximation}\label{sec:tuk}

Here is an excerpt  from taken form the introduction of \cite{TUK93B} entitled `Introduction to procedure orientation'.

\begin{quotation}
Davies has gone a long way toward procedure orientation. Her
discussion at the top of her page 29 is summarized by the sentence
(material in  [ ] added): ``In other words models are chosen to
produce [as properties of the procedures they `prescribe'] desirable
operational characteristics.'' 

This is a long step forward, models are now valued for their (formal)
consequences, rather than for their truth. This is good, but does not
go far enough. Since the formal consequences are consequences of the
truth of the model, once we have ceased to give a model's truth a
special role, we cannot allow it to ``prescribe'' a procedure. What we
really need to do is to choose a procedure, something we can be helped
in by a knowledge of the behavior of alternative procedures when
various models (= various challenges) apply, most helpfully models
that are (a) bland (see below) or otherwise relatively trustworthy,
(b) reasonable in themselves, and (c) compatible with the data. 

%\pagenumbering{arabic}
In short, we need to change from assumption-orientation to
procedure-orientation.\ 

The classical paradigm ran something like:\\
\setlength{\unitlength}{0.8cm}
\begin{picture}(17,3.5)
\put(0,1.3){\makebox(5,0.7)[l]{beliefs about the world}}
\put(0,0.25){\makebox(12.7,0.7)[l]{considerations about mathematical
    manageability (of optimization)}}
\put(12.75,1.3){\makebox(3,0.7)[l]{assumed model}}
\put(4.5,1.5){\vector(1,0){8}}
\put(12.25,0.73){\vector(1,2){0.3}}
\end{picture}

where, as Davies rightly emphasizes, the process involved in these
first two arrows has been very inadequately discussed, followed by\\ 
 
\setlength{\unitlength}{1cm}
\begin{picture}(17,3)
\put(3.2,2.2){\makebox(3,0.7)[l]{assumed model}}
\put(3.2,0.2){\makebox(4.7,0.7)[l]{criterion to be optimized}}
\put(8.6,1.5){\makebox(4,0.7)[l]{optimized procedure}}
\put(6.4,2.5){\vector(4,-1){2}}
\put(8,0.6){\vector(1,3){0.35}}
\end{picture}

beliefs about nature of world------------ assumed model\\
considerations about mathematical manageability (of optimization)\\

%assumed model--------------------\\
%                               ---------------------------------optimized procedure\\

The solidity of the optimization was classically taken as legitimizing
the unremovable fluidity of the choice as [sic] assumed model and of
the criterion to be optimized.

Since the model was about assumed (revealed?) truth, one could try to
get away with one model and with qualitative knowledge about
optimization -- what procedure optimized rather than how well it
performed in contrast to alternatives.

Once we have become procedure-oriented, we expect a very different
pattern of thought. Beginning somewhere, we enter a loop of
exploration and improvement:\\ 

\setlength{\unitlength}{0.8cm}
\begin{picture}(17,7)
\put(0.1,6){\parbox{11cm}{choose new procedure(s) in the light of
    available knowledge}}
\put(0.1,4){\parbox{5.5cm}{compare performance \newline of
    procedure-model pairs(*)}}
\put(11.5,4){\parbox{4cm}{discover what models challenge it effectively}}
\put(0.1,1.2){\parbox{11cm}{try these models, perhaps by
    simulation, on both the new procedure(s) and the leading old
    procedure(s)}}
\put(1.2,2){\vector(-2,3){0.8}}
\put(0.4,4.7){\vector(2,3){0.8}}
\put(11,6){\vector(1,-1){1.45}}
\put(12,3.5){\vector(-1,-3){0.5}}
 \end{picture}

choose new procedure(s)in the light of available knowledge\\

compare performance--------------------discover what models\\
of procedure-model pairs(*)--------------challenge it effectively\\

try these models, perhaps by simulation, on both the new\\
procedures(s) and the leading old procedure(s)

At (*) we are likely to seek some sort of saddle-point, where deviations of model move performance in one direction, while deviations of procedure move it in the other.

Once we have taken simulation seriously, the minimum mathematical manageability we require is limited to:\\
\begin{itemize}
\item being able to apply procedures to data, and\\
\item being able to simulate data from models\\
\end{itemize}
so that we no longer require an ability to carry out complex formal manipulations.\\

\begin{center}
*\qquad framework \qquad*
\end{center}
\addcontentsline{toc}{subsubsection}{\hspace*{0.455cm} framework}

We need to frame our thinking in terms of:\\
\begin{itemize}
\item questions, on which we hope the data - - and its analysis - - will shed light,\\
\item targets - - one or a few numbers or pictures (or both)- - to be calculated from the data by the use of \\
\item procedures - - well-specified computations leading from the data to values of numerical targets or examples of pictorial targets.
\end{itemize}
(The narrowness of targets reflects the limitations of human perception.)\\
\end{quotation}

Here is a second excerpt taken from Section 1  entitled `A is for Approximation'. It is also cited on page 24 of  \cite{DAV14}. 
\begin{quotation}
Davies's emphasis on approximation is well-chosen and surprisingly
novel. While these will undoubtedly be a place for much careful work
in learning how to describe the concept - - and its applications - -
in detail, it is clear that Davies has taken the decisive step by
asserting that there must be a formal admission that adequate
approximation, of one set of observable (or simulated) values by
another set, needs to be treated as practical identity. 

If, as is so convenient, we continue to use continuous models to describe - - or perhaps only to illuminate - - observed data, we should have to say that certain aspects of the data - - not typically, but unavoidably, including ``Most (modelled) observations have irrational values!''- - are not to be used in relating conceptual (or simulated) samples to observed samples. Thought and debate as to just which aspects are to be denied legitimacy will be both necessary and valuable.
\end{quotation}

It is exceptional for statisticians to be interested in approximation, most are wedded to the `assumed truth optimization paradigm' described above. One exception is  \cite{COXBATT17} who point out that in high dimensions a single model does not reflect the complexity of the situation. They write
\begin{quotation}
Data with a small number of study individuals and a large number of potential explanatory features arise particularly in genomics. Existing methods of analysis result in a single model, but other sparse choices of explanatory features may fit virtually equally well.
\end{quotation}

There are several forms of asymptotics. There are asymptotics which work for the sample size $n_0$ of the data at hand, there are asymptotics which work for some known $n_1>n_0$ and there are asymptotics which work for some unknown $n_1$.  For data analysis only the first form is of interest. Here is Tukey on asymptotics
\begin{quotation}
Davies makes much of the discontinuity of asymptotics in the topology of classical inference. Here [sic] remarks are accurate, her conclusions unhelpful. This is another~free lunch~ issue. There could be asymptotic free lunches, but no one seems demonstrably to have ever had one.If asymptotics are of any real value, it must be because they teach us something useful in finite samples. I wish I knew how to be sure when this happens.But leaving this basic doubt aside for the present, classical asymptotics (like those cases where they do their best) should be read as saying:\newline
1) Things might not be any better than so-and-so for large samples.\newline
2) No one has a real example where they are better.\newline
3) Thus they give both lower bounds and reasonable choices for an appropriate diffidence (= lack of confidence).\newline
We need to do asymptotics based on bland models - - how lucky we are that the Gaussian is relatively bland.
\end{quotation}

Such results are required as they allow valid post-model selection inference, see for the problems of inference after model selection.  54 approximations  with a sum of squared residuals of less than 9. 

\section{Acknowledgment} 
The theory of Gaussian P-values central to this paper was developed together with Lutz D\"umbgen. Joe Whittaker gave a proof of based on Cochran's theorem and found many errors in the R package {\it gausscov}..  
%\begin{thebibliography}{}
\bibliographystyle{apalike}
%\bibitem{plainnat}
\bibliography{literature}

\begin{thebibliography}{}

\bibitem[Benjamini and Hochberg, 1995]{BENJHOCH95}
Benjamini, Y. and Hochberg, Y. (1995).
\newblock Controlling the false discovery rate: A practical and powerful
  approach to multiple testing.
\newblock {\em Journal of the Royal Statistical Society: Series B},
  57:289--300.

\bibitem[Brillinger, 2002]{BRIL02A}
Brillinger, D.~R. (2002).
\newblock John {W}. {T}ukey: the life and professional contributions.
\newblock {\em Annals of Statistics}, 30:1535--1575.

\bibitem[B\"uhlmann et~al., 2014]{BKL14}
B\"uhlmann, P., Kalisch, M., and Meier, L. (2014).
\newblock High-dimensional statistics with a view toward applications in
  biology.
\newblock {\em Annual Reviewof Statistics and Its Application}, 1(1):255--278.

\bibitem[Cox and Battey, 2017]{COXBATT17}
Cox, D.~R. and Battey, H.~S. (2017).
\newblock Large numbers of explanatory variables, a semi-descriptive analysis.
\newblock {\em Proc. Natl. Acad. Sci. USA}, 114(32):8592–--8595.

\bibitem[Davies, 2014]{DAV14}
Davies, L. (2014).
\newblock {\em Data Analysis and Approximate Models}.
\newblock Monographs on Statistics and Applied Probability 133. CRC Press.

\bibitem[Davies, 2018]{DAV18a}
Davies, L. (2018).
\newblock On {P}-values.
\newblock {\em Statistica Sinica}, 28:2823--2840.

\bibitem[Davies and D\"umbgen, 2022]{DAVDUEM22}
Davies, L. and D\"umbgen, L. (2022).
\newblock Covariate selection based on a model-free approach to linear
  regression with exact probabilities.
\newblock arxiv.org/abs/2202.01553v2.

\bibitem[Davies, 1995]{DAV95}
Davies, P.~L. (1995).
\newblock Data features.
\newblock {\em Statistica Neerlandica}, 49:185--245.

\bibitem[Davies, 2008]{DAV08}
Davies, P.~L. (2008).
\newblock Approximating data (with discussion).
\newblock {\em Journal of the Korean Statistical Society}, 37:191--240.

\bibitem[Davies and Kovac, 2001]{DAVKOV01}
Davies, P.~L. and Kovac, A. (2001).
\newblock Local extremes, runs, strings and multiresolution (with discussion).
\newblock {\em Annals of Statistics}, 29(1):1--65.

\bibitem[Dezeure et~al., 2015]{DEBUEMEME15}
Dezeure, R., B\"uhlmann, P., Meier, L., and Meinshausen, N. (2015).
\newblock High-dimensional inference: confidence intervals, p-values and
  {R}-software hdi.
\newblock {\em Statistical Science}, 30(4):533--558.

\bibitem[D\"umbgen and Davies, 2023]{DUMDAV23}
D\"umbgen, L. and Davies, L. (2023).
\newblock Connecting model-based and model-free approaches to linear least
  squares regression.
\newblock arxiv.org/abs/1807.09633.

\bibitem[Feng et~al., 2020]{FFSS20}
Feng, Y., Fan, J., Saldana, D., and Samworth, R. (2020).
\newblock Sis: Sure independence screening.
\newblock https://CRAN.R-project.org/package=SIS.

\bibitem[Friedman et~al., 2019]{FRHATI19}
Friedman, J., Hastie, T., and Tibshirani, R. (2019).
\newblock Graphical lasso: Estimation of gaussian graphical models.
\newblock https://CRAN.R-project.org/package=glasso.

\bibitem[Golub et~al., 1999]{GOLETAL99}
Golub, T., Slonim, D., P., T., Huard, C., Gaasenbeek, M., Mesirov, J., Coller,
  H., Loh, M., Downing, J., Caligiuri, M., Bloomfield, C., and Lander, E.
  (1999).
\newblock Molecular classification of cancer: class discovery and class
  prediction by gene expression monitoring.
\newblock {\em Science}, 286(15):531--537.

\bibitem[Hampel, 1985]{HAM85}
Hampel, F.~R. (1985).
\newblock The breakdown points of the mean combined with some rejection rules.
\newblock {\em Technometrics}, 27:95--107.

\bibitem[Hastie et~al., 2020]{HASTIBTIB20}
Hastie, T., Tibshiranie, R., and Tibshirani, R. (2020).
\newblock Best subset, forward stepwise or lasso? analysis and recommendations
  based on extensive comparisons.
\newblock {\em Statistical Science}, 35(4):579--592.

\bibitem[Huber, 2011]{HUB11}
Huber, P.~J. (2011).
\newblock {\em Data Analysis}.
\newblock Wiley, New Jersey.

\bibitem[Jiang et~al., 2021]{JFLRLWLZ21}
Jiang, H., Fei, X., Liu, H., Roeder, K., Lafferty, J., Wasserman, L., Li, X.,
  and Zhao, T. (2021).
\newblock huge: High-dimensional undirected graph estimation.
\newblock https://CRAN.R-project.org/package=huge.

\bibitem[Kuchibhotla et~al., 2022]{KKK22}
Kuchibhotla, A., Kolassa, J., and Kuffnr, T. .~J. (2022).
\newblock Post-selection inference.
\newblock {\em Annual Review of Statistics and Its Application}, 9:505--527.

\bibitem[Laszkiewicz, 2021]{LASZ21}
Laszkiewicz, M. (2021).
\newblock Thresholded adaptive validation: Tuning the graphical lasso for graph
  recovery.
\newblock https://github.com/MikeLasz/thav.glasso.

\bibitem[Lockhart, 2017]{LOCK17}
Lockhart, R. (2017).
\newblock Inference in high-dimensional linear models course notes.
\newblock httpDimensionals://www.sfu.ca/~lockhart/richard/Cambridge/Notes.pdf.

\bibitem[Meier et~al., 2021]{MDMMB21}
Meier, L., Dezeure, R., Meinshausen, N., Maechler, M., and Buehlmann, P.
  (2021).
\newblock hdi: High-dimensional inference.
\newblock https://CRAN.R-project.org/package=hdi.

\bibitem[Meinshausen and B\"uhlmann, 2006]{MEIBUE06}
Meinshausen, N. and B\"uhlmann, P. (2006).
\newblock High-dimensional graphs and variable selection with the lasso.
\newblock {\em Annals of Statistics}, 34(3):1436--1462.

\bibitem[Scott, 2022]{SCO22}
Scott, S. (2022).
\newblock Boomspikeslab: {MCMC} for spike and slab regression.
\newblock https://cran.r-project.org/web/packages/BoomSpikeSlab/index.html.

\bibitem[Seheult and Tukey, 2001]{SEHTUK01}
Seheult, A.~H. and Tukey, J.~W. (2001).
\newblock Towards robust analysis of variance.
\newblock In Saleh, A.~K.~M.~E., editor, {\em Data Analysis from Statistical
  Foundations: A Festschrift in Honor of the 75th Birthday of D.A.S. Fraser},
  pages 217--244. Nova Science, New York.

\bibitem[Sun, 2019]{SUN19}
Sun, T. (2019).
\newblock Scaled sparse linear regression.
\newblock https://CRAN.R-project.org/package=scalreg.

\bibitem[Tukey, 1993a]{TUK93D}
Tukey, J.~W. (1993a).
\newblock Discussion- {D}avies's data sets.
\newblock Princeton University, Princeton.

\bibitem[Tukey, 1993b]{TUK93C}
Tukey, J.~W. (1993b).
\newblock How {D}avies's data sets might reasonably be approached.
\newblock Princeton University, Princeton.

\bibitem[Tukey, 1993c]{TUK93B}
Tukey, J.~W. (1993c).
\newblock Issues relevant to an honest account of data-based inference,
  partially in the light of {L}aurie {D}avies's paper.
\newblock Princeton University, Princeton.

\bibitem[von Neumann, 1947]{wrks47}
von Neumann, J. (1947).
\newblock The {M}athematician.
\newblock In Adler, M. and Heywood, R., editors, {\em The Works of the Mind},
  volume~1, pages 180--196. University of Chicago Press, Chicago.

\end{thebibliography}

\section{R procedures}
\subsection{Simulating Gaussian P-values}\label{R0}
{\footnotesize
\begin{verbatim}
>library(gausscov)
> data(snspt)
> a<-ar.mle(snspt)   # estimate lag using mle and AIC,
> a[[1]]                    # fails as always  
[1] 12
> snsptl<-flag(snspt,3253,1,12,inr=F)  # calculate the lags using the function flag
> b<-lm(snsptl[[1]]~snsptl[[2]])    # autoregression
> summary(b)                    # lag 7 has an F P-value of 0.566
> simgpval(snsptl[[1]],snsptl[[2]],7,1000)  # run the simulation 1000 times, 
                   # Gaussian P-values of the Gaussian covariates plotted, 
                   # they are clearly uniformly distributed over (0,1)
[[1]]
[1] 0.5660791 0.5820000 # F -Pvalue 0.566 as above and relative 
                        # frequency of Gaussian covariates being better
\end{verbatim}
}

\subsection{Table~\ref{tab:ribo1}}\label{R2}
{\footnotesize
\begin{verbatim}
library(gausscov)
library(hdi)
data(riboflavin)
t<-system.time(a<-f3st(riboflavin[[1]],riboflavin[[2]],m=1,kmn=15))[[1]]  
print(t)   # takes about 1.5 seconds 
a[[1]]
 b<-lm(riboflavin[[1]]~riboflavin[[2]][,a[[1]][1,2:6]])   
summary(b)
plot(b$res)
plot(riboflavin[[1]])
lines(riboflavin[[1]]-b$res,col='red')   # the approximation
t<-system.time(a<-f3st(riboflavin[[1]],riboflavin[[2]],m=3,kmn=15))[[1]]  # takes about 25 seconds
a[[1]]
 b<-lm(riboflavin[[1]]~riboflavin[[2]][,a[[1]][1,2:10]])   
summary(b)
plot(b$res)
plot(riboflavin[[1]])
lines(riboflavin[[1]]-b$res,col='red')   # the approximation
\end{verbatim}
}

\subsection{Table~\ref{tab:vardat1}}\label{R1}
{\footnotesize
\begin{verbatim}
library(BVAR) 
library(coda)
data<-fred_qd  # original USA economic data
dim(data)
library(gausscov)
data(vardata)    # USA economic data without NAs
lvardata1<-flag(vardata,256,1,10,T) # select variable 1 USA GDP,  lags 1:10 of all covariates
t<-system.time(a<-f3st(lvardata1[[1]],lvardata1[[2]],m=1,kmn=16))[[1]]
t              # takes about 26 seconds
a[[1]]  # columns in order (1) sums of squared residuals  (2) selected covariates
 a[[1]][1:2,]
b<-lm(lvardata1[[1]]~lvardata1[[2]][,a[[1]][1,2:12]])
summary(b)
res<-b$res 
acf(res)  # a slight degree of autocorrelation, will be ignored.$
sd(re
\end{verbatim}
}

\subsection{The leukemia data}\label{R1.5}
{\footnotesize
\begin{verbatim}
data(leukemia)
 a<-f1st(leukemia[,3572],leukemia[,1:3571])
a[[1]]
plot(leukemia[,3572])
points(leukemia[,3572]-a[[2]], col=''red'')
 a<-f1st(leukemia[,3572],leukemia[,1:3571],kmn=10)
a[[1]]
plot(leukemia[,3572])
points(leukemia[,3572]-a[[2]], col=''red'')
a<-f3st((leukemia[,3572],leukemia[,1:3571],m=4,kmn=15)
a[[1]][1:2,]
\end{verbatim}
}

\subsection{Boston data}\label{R31}
{\footnotesize
\begin{verbatim}
> abst<-f1st(boston[,14],boston[,1:13])
> sum(abst[[2]]^2)
[1] 11868.24
> abst[[1]]
     [,1]          [,2]         [,3]         [,4]
[1,]   13  -0.537153086 2.716050e-25 3.880071e-26
[2,]    6   4.294369137 8.084661e-23 1.154952e-23
[3,]   11  -0.973694912 3.374442e-16 4.820631e-17
[4,]    8  -1.123472062 2.048889e-10 2.926984e-11
[5,]    5 -16.677063917 2.930785e-06 4.186841e-07
[6,]    4   3.051944251 3.756968e-03 5.375760e-04
[7,]   12   0.008977955 5.391123e-03 7.719459e-04
[8,]    0  30.411960721 1.187252e-09 1.187252e-09
\end{verbatim}
}

\subsection{Table~\ref{tab:bost1}}\label{R3}
{\footnotesize
\begin{verbatim}
library(gausscov)
data(boston)
bostint<-fgeninter(boston[,1:13],8)  # generate the interactions
t<-system.time(a<-f3st(boston[,14],bostint[[1]]),m=1,kmn=15)[[1]]  
print(t)
a[[1]]
a[[1]][1:2,]
 b<-lm(boston[,14]~bostint[,a[[1]][1,2:12]])   
summary(b)
plot(b$res) # the residuals
plot(boston[,14])
lines(boston[,14]-b$res,col=''red'')   # the approximation
bostint[[2]][191110,]  #    interaction 191110
\end{verbatim}
}

\subsection{Sunspot data: autoregression}\label{R41}
{\footnotesize
\begin{verbatim}
data(snspt)
 length(snspt)
lsnspt<-flag(snspt,3253,1,150) 
a<-f1st(lsnspt[[1]],lsnspt[[2]])
 a[[1]]
     [,1]        [,2]          [,3]          [,4]
[1,]    1  0.54073955 5.602675e-178 3.917955e-180
[2,]    2  0.12443321  4.249741e-09  2.971847e-11
[3,]    4  0.12976551  5.503330e-12  3.848483e-14
[4,]    6  0.07657507  6.300486e-04  4.407313e-06
[5,]    9  0.11262754  1.987266e-11  1.389697e-13
[6,]   27 -0.08019322  3.861438e-18  2.700306e-20
[7,]  111  0.05069863  9.056070e-09  6.332916e-11
[8,]    0  3.67318330  1.092745e-04  1.092745e-04
res<-a[[2]]
sd(res)
plot(res)  
tres<-res/(lsnspt[[1]]+10)
acf(tres)
ltres<-flag(tres,3103,1,10)
b<-f1st(ltres[[1]],ltres[[2]])
b[[1]]
     [,1]        [,2]         [,3]         [,4]
[1,]    5  0.13627475 1.367908e-13 2.735816e-14
[2,]    3  0.13401146 2.105358e-13 4.210716e-14
[3,]    8  0.10094316 8.327526e-08 1.665505e-08
[4,]    7  0.07158877 3.125446e-04 6.251674e-05
[5,]    4  0.07292715 1.817183e-04 3.634631e-05
[6,]    2  0.06011486 3.751362e-03 7.514007e-04
[7,]    0 -0.05419739 3.825415e-11 3.825415e-11
delta<-b[[2]]
acf(delta)
sdel<-sort(delta)
 plot(pt(2*(pmin(sdel,0)/0.6),2)-0.5+pt(2*(pmax(sdel,0)/0.5),10),1:3093/3093,t="l")
\end{verbatim}
}

\subsection{Sunspot data: non-parametric regression}\label{R42}
{\footnotesize
\begin{verbatim}
sntrg<-fgentrig(3253,1626)  # generate sines and cosines
a<-f1st(snspt,sntrg[[1]],p0=0.001)
res<-a[[2]]
snf<-snspt-res  # approximation
 plot(snf,t="l") # plot approximation,  some values negative
snff<-20*exp(log(1+snf/20))
plot(snff, t=''l'')
res<-snspt-snff
acf(res)         #residuals correlated
lres<-flag(res,3253,1,10)  
b<-f1st(lres[[1]],lres[[2]],10)   # autregression
rres<-b[[2]]
plot(rres)  # outliers
length(rres)
ind<-(1:3243)[abs(res-median(res))<3.5*mad(res)]  # remove outliers
length(ind)
acf(rres[ind])
length(ind)
plot(pt(sort(rres[ind])/16,3),(1:3159)/3159)
\end{verbatim}
}

%\newpage
\section{Figures}
\begin{figure}[hb]
\title{Riboflavin data: approximation and residuals}
\includegraphics[width=1\textwidth,height=160px]{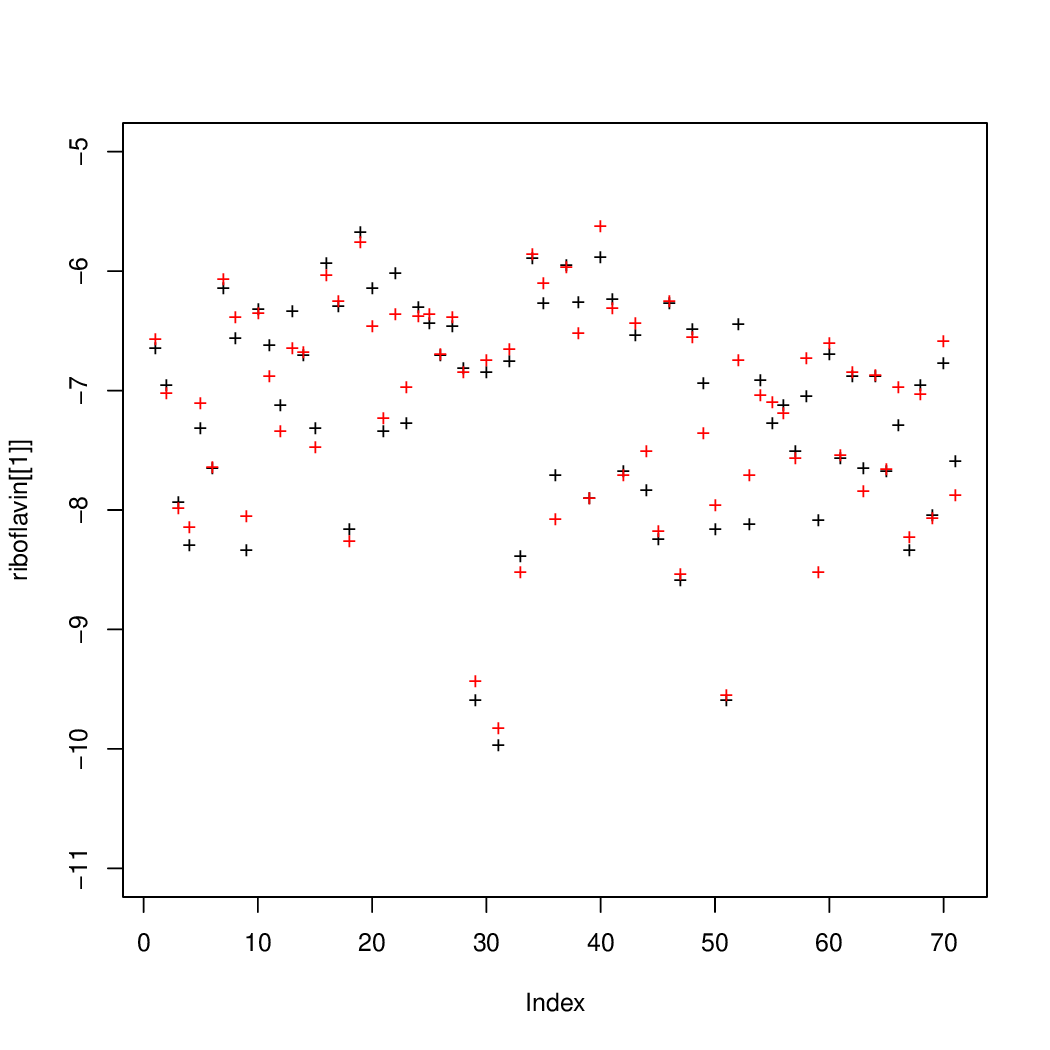}
\includegraphics[width=1\textwidth,height=160px]{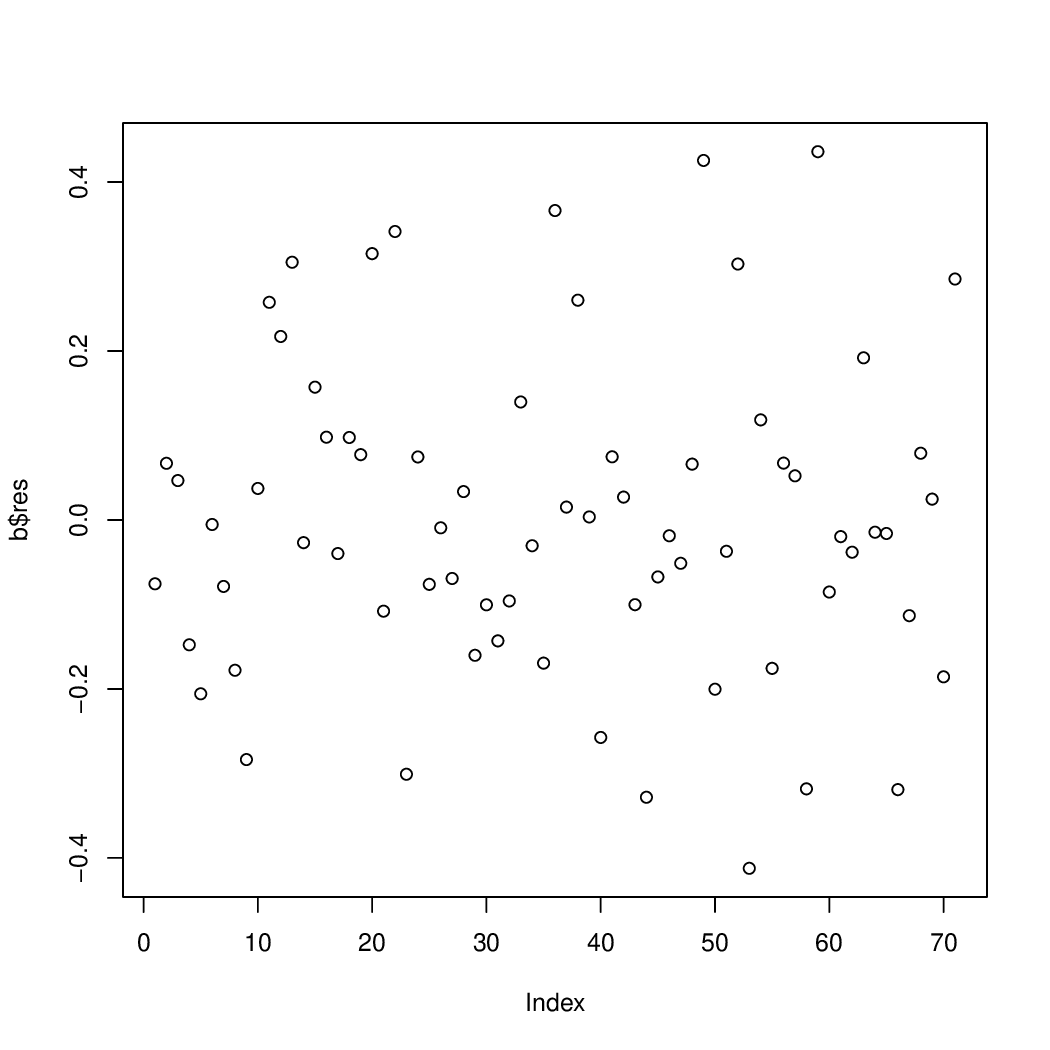}
\caption{Top: data and the approximation (red). Bottom: the residuals. \label{fig:riboflavin}}
\end{figure}

\begin{figure}[hb]
\title{Riboflavin data: approximation and residuals}
\includegraphics[width=1\textwidth,height=160px]{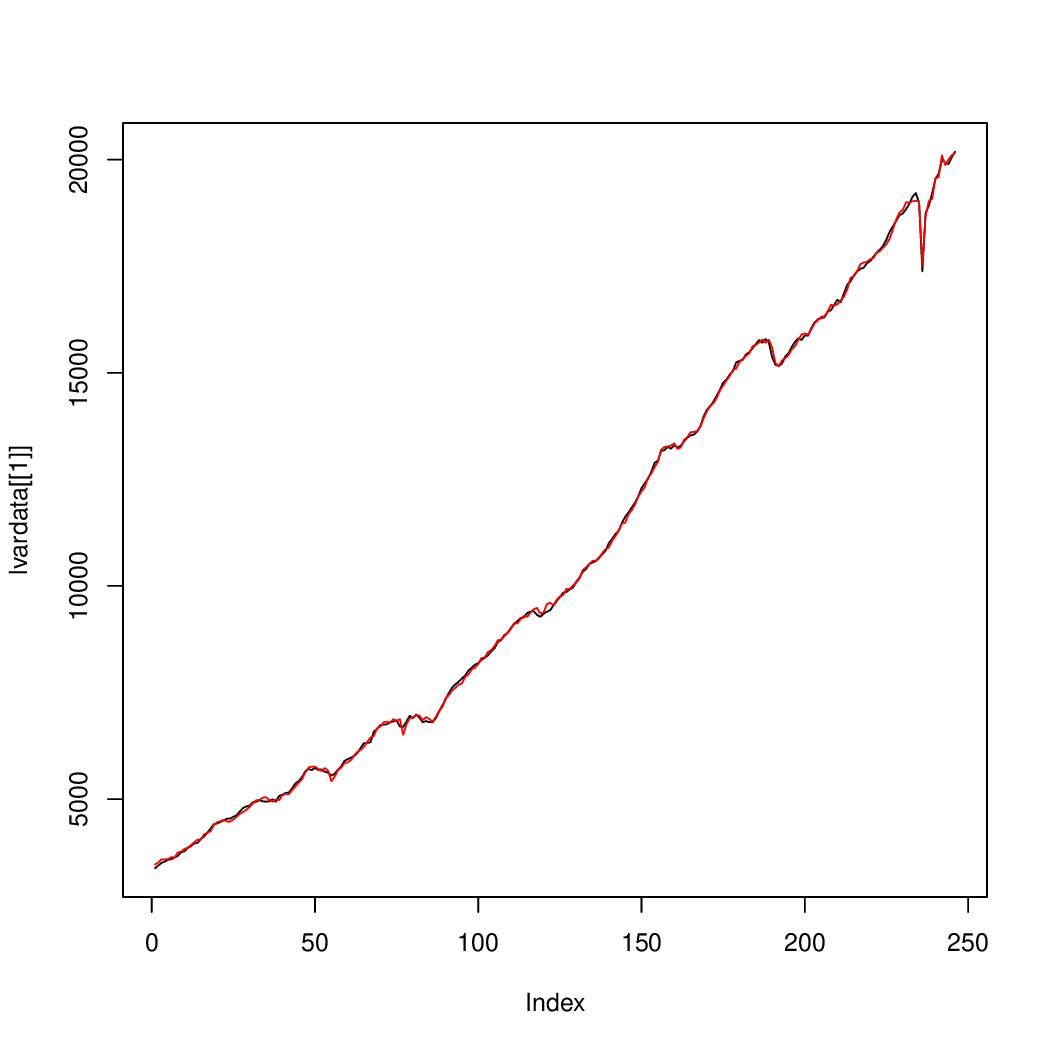}
\includegraphics[width=1\textwidth,height=160px]{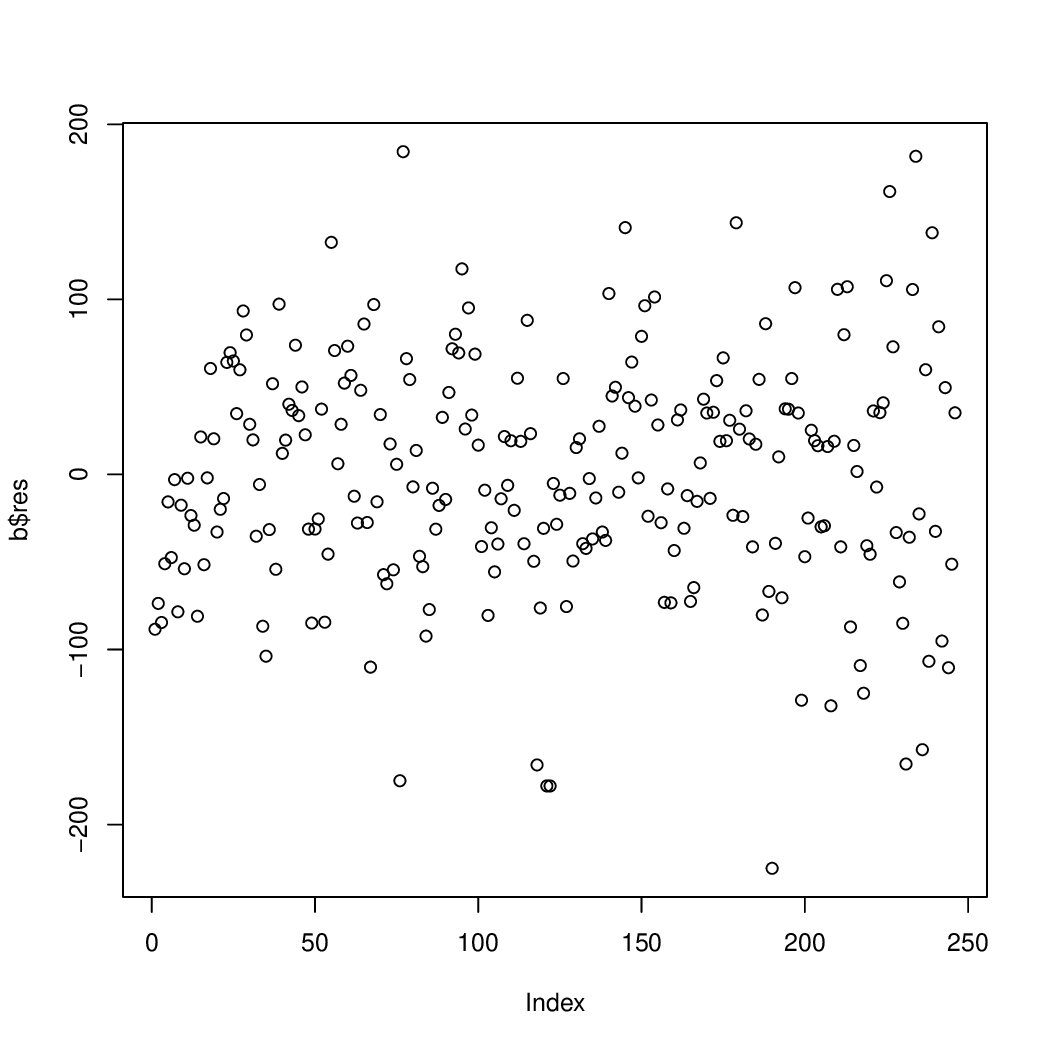}
\caption{Top: USA GDP and the approximation (red). Bottom: the residuals. \label{fig:vardata}}
\end{figure}

\begin{figure}[hb]
\title{Model for the sunspot data}
\includegraphics[width=1\textwidth,height=150px]{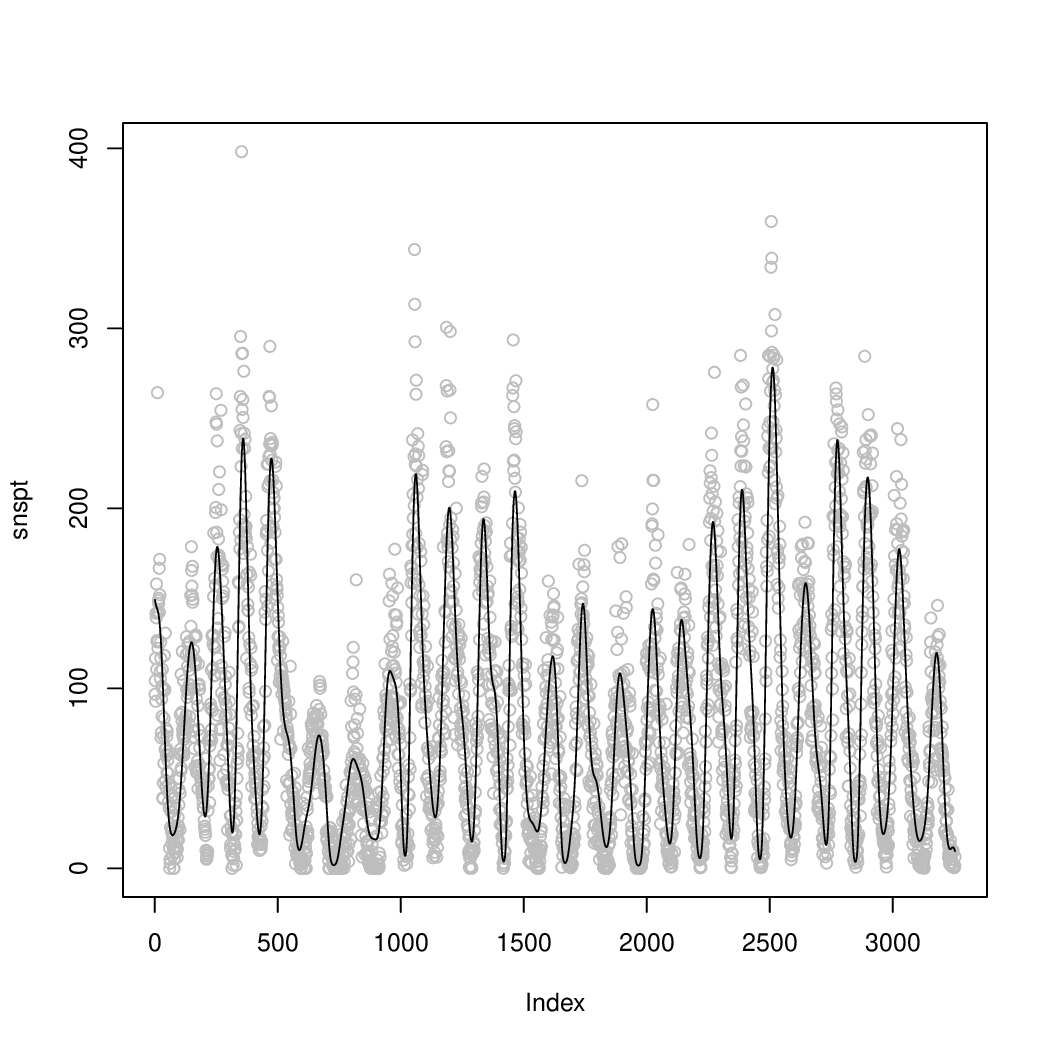}
\includegraphics[width=1\textwidth,height=150px]{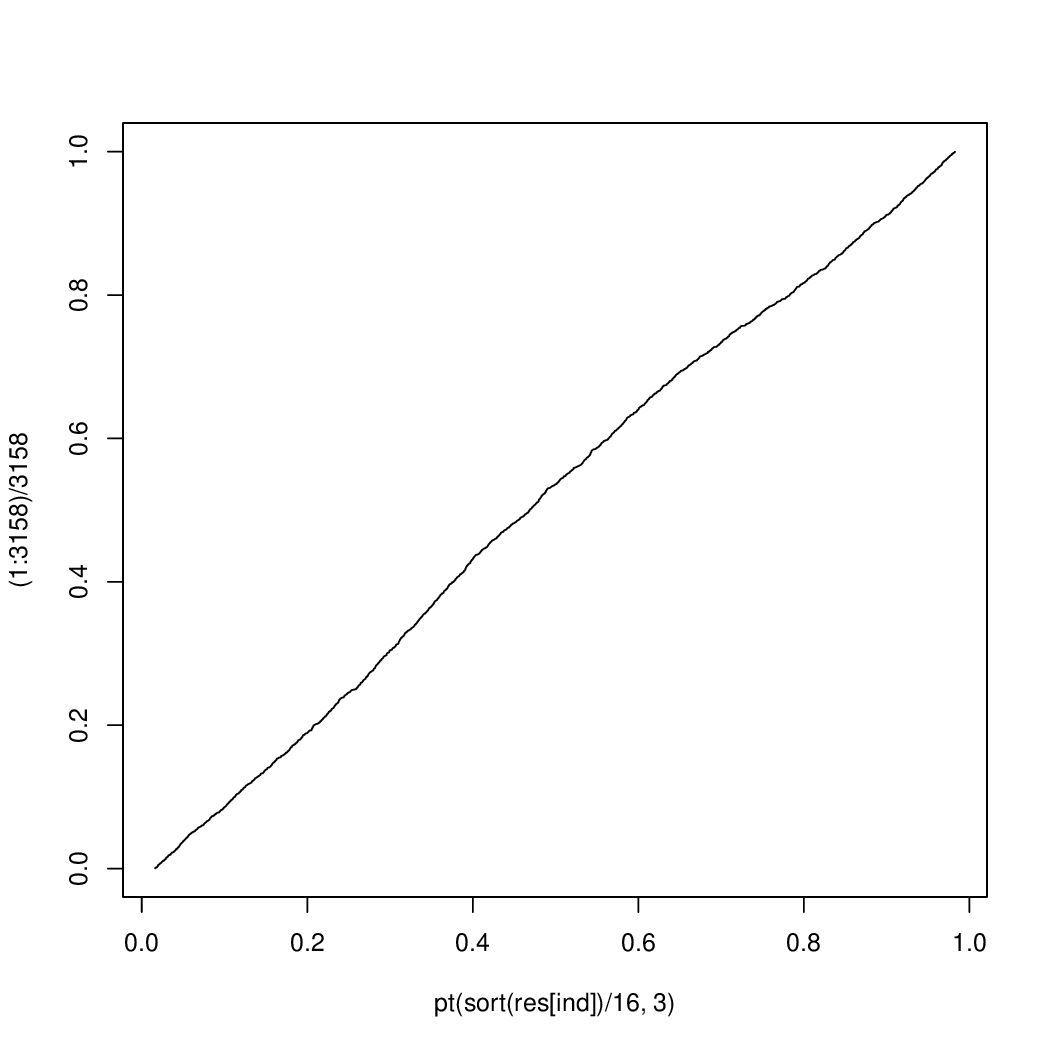}
\caption{Top:data and the non-parametric regression function. Bottom: empirical distribution function of autoregessive error term divided by 16 with $F=t_3$. \label{fig:snspt_mod}}
\end{figure}

\begin{figure}[t]
\title{-log(gpvalues)}
\includegraphics[width=1\textwidth,height=250px]{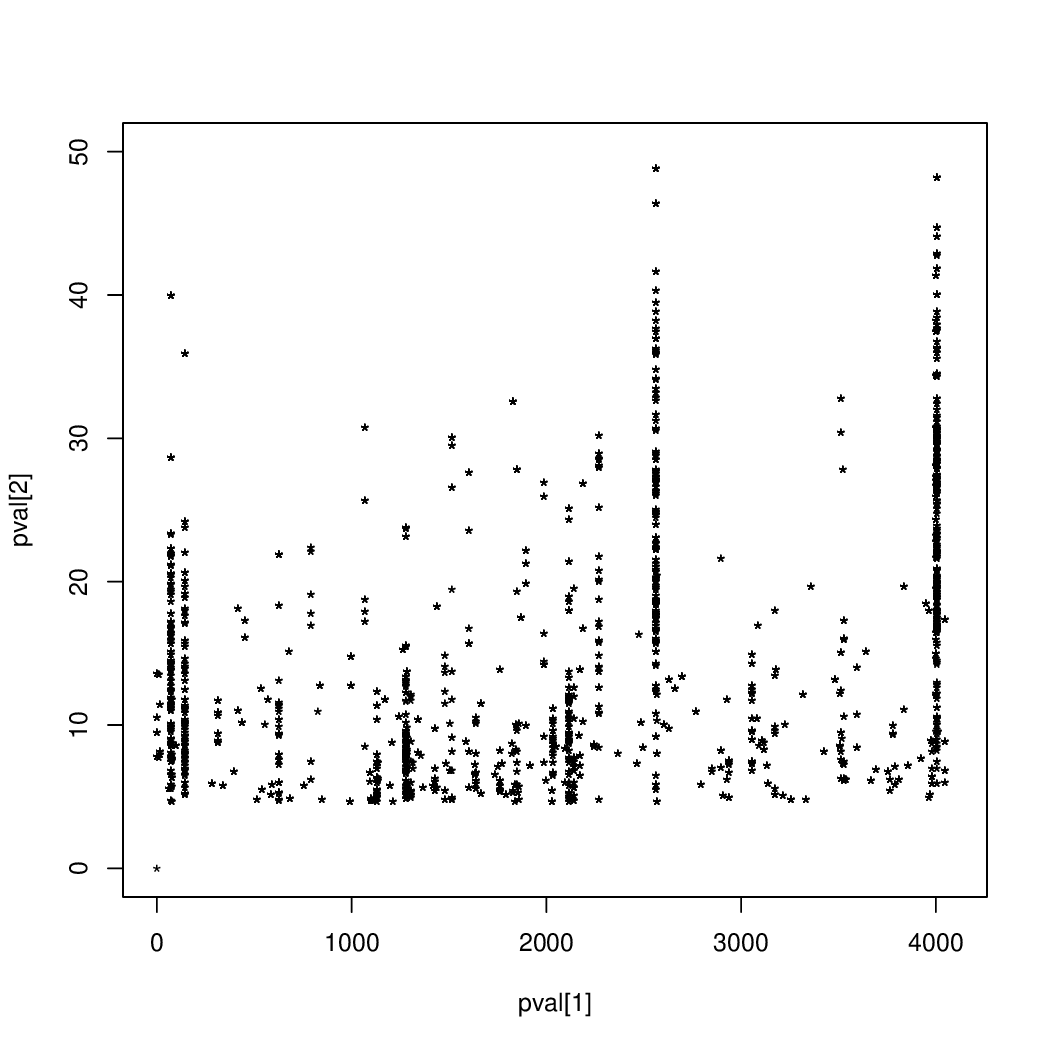}
\caption{Plot of $-log(gpvalue)$ for each of the 190 covariates in the approximations to the riboflavin data. \label{fig:log_pval_ribo}}
\end{figure}

\begin{figure}[t]
\title{Dependency graph}
\includegraphics[width=1\textwidth,height=250px]{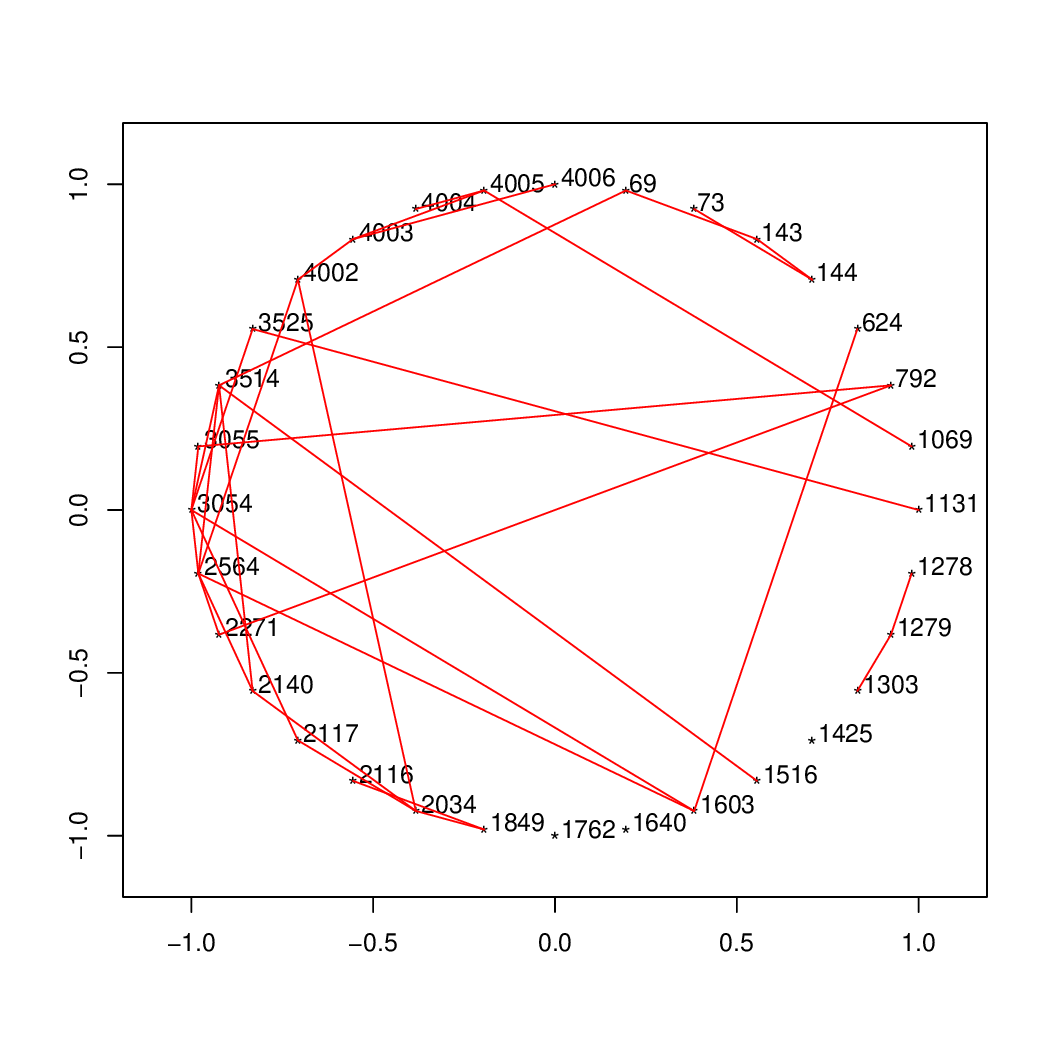}
\caption{This is an undirected dependency graph for the 32 covariates of Table~\ref{tab:frequencies}. \label{fig:dep_graph}}
\end{figure}

\begin{figure}[t]
\includegraphics[width=1\textwidth,height=250px]{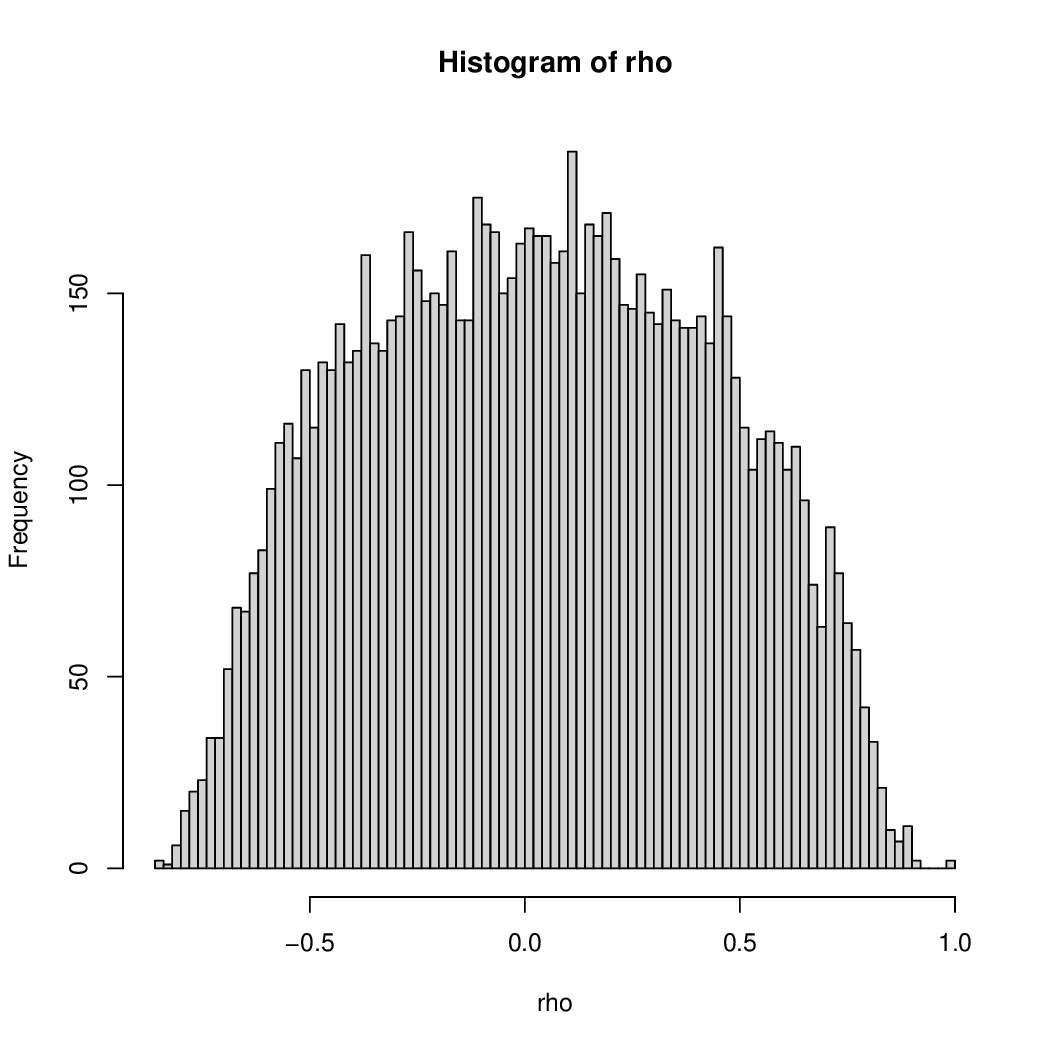}
\caption{Histogram of the correlations of the covariates of the riboflavin data. \label{fig:ribo_cor}}
\end{figure}

\end{document}